\documentclass[letterpaper,twocolumn,showpacs,epl,superscriptaddress]{revtex4}
\usepackage[latin1]{inputenc}
\usepackage{bm}
\usepackage{multirow}
\usepackage{amssymb}
\usepackage{amsbsy}
\usepackage{amsmath}
\usepackage{stmaryrd}
\usepackage{graphicx}
\usepackage{subfigure}
\usepackage{epsfig}
\usepackage{hyperref}
\makeatletter
\usepackage{pifont}
\makeatother

\begin{document}

\title{Transient fluctuation theorem in closed quantum systems}

\author{Christian Bartsch}
\email{cbartsch@uos.de}
\affiliation{Fachbereich Physik, Universit\"at Osnabr\"uck,
             Barbarastrasse 7, D-49069 Osnabr\"uck, Germany}

\author{Jochen Gemmer}
\email{jgemmer@uos.de}
\affiliation{Fachbereich Physik, Universit\"at Osnabr\"uck,
             Barbarastrasse 7, D-49069 Osnabr\"uck, Germany}

\date{\today}

\begin{abstract}

Our point of departure are the unitary  dynamics of closed quantum systems as generated from the Schr\"odinger equation. We focus on a class of quantum models that typically exhibit roughly exponential relaxation of some observable within this framework. Furthermore, we focus on pure state evolutions. An entropy in accord with Jaynes principle is defined on the basis of the quantum expectation value of the above observable. It is demonstrated that the resulting deterministic entropy dynamics are in a sense in accord with a transient fluctuation theorem. Moreover, we demonstrate that the dynamics of the expectation value are describable in terms of an Ornstein-Uhlenbeck process. These findings are demonstrated numerically and supported by analytical considerations based on quantum typicality.

\end{abstract}


\maketitle

\textit{1. Introduction}
Fluctuation theorems as general rules controlling the entropy production in all sorts of physical systems have been in the focus of nonequilibrium physics for roughly two decades \cite{evans1993,evans1994,gallavotti1995,seifert2005}. Especially fluctuation theorems describing distributions of work have been extensively analyzed for classical deterministic and stochastic dynamics \cite{jarzynski1997,crooks1999,engel2009}. In the context of quantum mechanics most investigations in this field are also on quantum work fluctuation relations \cite{mukamel2003,crooks2008,campisi2009,talkner2008}. These approaches usually require some notion of work performed by a time-dependent Hamiltonian and the system to be initially in a canonical Gibbs state \cite{campisi2011}. In this Letter in contrary we do not consider any work but fluctuations of entropy in a class of closed non-driven quantum systems. Furthermore, we focus on pure states rather than Gibbs states. The entropy will be given
  below simply as a function of the (pure) state of the system. Thus ``entropy fluctuations'' here refer to deterministic but irregular appearing (small) deviations of the entropy evolution from a smooth mean behavior. To some extent comparable scenarios have been 
studied in \cite{esposito2006,esposito2007-1,esposito2009,esposito2010}. 
However, none of these works considers the specific entropy definition used in this paper in combination with the time evolution of pure states directly arising from the Schr\"odinger equation.

In general the fluctuation theorem (FT) is said to hold if the probability, or in the case at hand rather the relative frequency, of mean entropy productions (averaged over a time step $\tau$) $\Sigma_{\tau}$, obeys the following relation:
\begin{equation}
\frac{P(\Sigma_{\tau})}{P(-\Sigma_{\tau})} = e^{\Sigma_{\tau}\cdot \tau} \ .
\label{fluc}
\end{equation}
We analyze relaxation processes in non-driven systems, thus we do not allude to time reversed trajectories as often done in this context, i.e., here (\ref{fluc}) describes the probabilities of producing/annihilating certain amounts of entropy along a time-evolving trajectory.
Moreover Eq. (\ref{fluc}) is meant as a transient FT
(\ref{fluc}) is meant to apply for time steps $\tau$ that are short compared to the time scale of the relaxation dynamics ($\tau \ll \tau_R$).

The probability density function does not necessarily have to be Gaussian to fulfill the FT. 
But if it is Gaussian which is (approximately) the case in our numerical calculations (see below), it has to take on the following form in order to fulfill the FT:
\begin{equation}
P(\Sigma_{\tau})= \frac{1}{\sqrt{2\pi}\sigma} e^{-\frac{1}{2}\frac{(\Sigma_{\tau}-\Sigma_{0})^{2}}{\sigma ^{2}}} , \qquad \sigma ^{2}=\frac{2\Sigma_{0}}{\tau}\ ,
\label{gauss}
\end{equation}
with $\Sigma_{0}$ being the mean entropy production and $\sigma ^{2}$ the variance.

\begin{figure}[htb]
\centering
\hspace{-1.0cm}
\includegraphics[width=6.4cm]{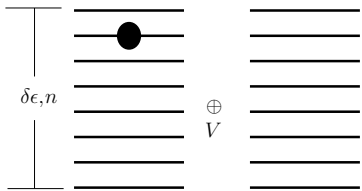}
\hspace{1.0cm}
\caption{Sketch of ``quantum two site hopping model''. The model consists of two energy bands, both of bandwidth $\delta\epsilon$, each containing $n$ levels with equidistant level spacing. Transitions between both bands are triggered by the ``perturbation'' $V$.} 
\label{model}
\end{figure}

\textit{2. Introduction of the Model and Expectation Value Dynamics}
Our numerics are based on a model which is designed to represent the most simple closed quantum model featuring exponential relaxation of an expectation value (a similar model and its overall behavior have already been established in \cite{bartsch2008}). The model, which is sketched in Fig. \ref{model}, describes a ``quantum two site hopping model''.  All operators are given on the level of discrete finite matrices.
The eigenvalues of some ``unperturbed Hamiltonian'' $H_0$ form two ''bands``, i.e., a left and a right band, both of width $\delta\epsilon$, both with equidistant level spacing. 
Furthermore, there is a ''perturbation`` $V$ consisting of transition operators, representing transitions between the two bands. The full Hamiltonian $H=H_0 + V$ reads
\begin{eqnarray}
H_0 &=& \sum_{i = 0}^{n-1} \frac{i}{n-1} \, \delta \epsilon \; | i, L
\rangle \langle i, L | + \sum_{j = 0}^{n-1} \frac{j}{n-1} \, \delta
\epsilon \; | j, R \rangle \langle j, R | \ , \nonumber \\
V&=& \left( \sum_{i,j = 0}^{n-1} v_{ij} \; | i, L \rangle \langle j,R | +
\text{H.c.} \right),\lambda^2 = \frac{1}{n^2} \sum_{i,j = 0}^{n-1} | v_{ij} |^{2}, \nonumber \\
\end{eqnarray}
where the $v_{ij}$ are chosen as random complex Gaussian numbers and $\lambda$ measures the perturbation strength. 
Note that the $v_{ij}$ are randomly chosen but fixed numbers in this definition. This applies for all the numerics following further below.
$| i, L \rangle$ ($| j, R \rangle$) are the eigenstates of $H_0$ which form the left (right) band. 
We only consider one relevant observable $A$ having two eigenvalues, $1$ and $-1$, each $n$-fold degenerate. Each eigenstate of $A$ also corresponds to an eigenstate of $H_0$. The $1$-eigenspace coincides with the left band, the $-1$-eigenspace with the right band. If the expectation value of $A$ is $1$ ($-1$), the excitation probability is completely concentrated on the left (right) band and if the expectation value is $0$, the excitation probability is equally partitioned between both bands. One may now define the projectors on the states forming the left (right) band as $P_L$ ($P_R$). Then $A$ may also be expressed as $A=P_L -P_R$ and therefore be interpreted as a ``position observable'' which measures the ``occupation asymmetry'' between the two bands.

\begin{figure}[htb]
\centering
\hspace{-1.0cm}
\includegraphics[width=6.4cm]{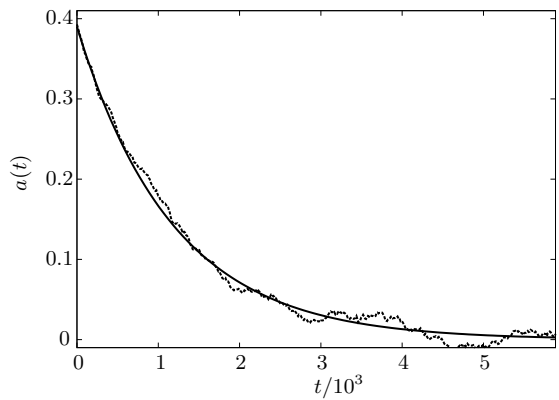}
\hspace{1.0cm}
\caption{Dynamics of expectation value $a(t)$. ``Fluctuating'' curve as generated from the Schr\"odinger Equation (dashed line) versus mean exponential behavior with the decay rate $R$ (solid line). System parameters: $N=6000$, $\delta\epsilon = 0.75$, $\lambda= 0.00013$.} 
\label{expcurve}
\end{figure}

In the following we analyze the dynamics of the expectation value $a(t)=\langle \psi(t) \vert A \vert \psi(t) \rangle$ for pure states. Thereby we focus on situations, where the rough system parameters $\delta\epsilon , n, \lambda$ are chosen from a suitable parameter range, such that the dynamics of $a(t)$, as generated by the full time-dependent Schr\"odinger Equation, result approximately as an exponential decay.
The suitable parameter range is essentially defined by
\begin{equation}
\frac{16 \pi^{2} \, n \, \lambda^{2}}{\delta \epsilon^{2}} \ll 1
\, , \quad \frac{8 \pi^{2} \, n^{2} \, \lambda^{2}}{\delta
\epsilon^{2}} > 1 \, . \label{eq-crit}
\end{equation}
Eq. (\ref{eq-crit}) implies that relaxation times $\tau_R$ have to be much longer than correlation times $\tau_C$, i.e., $\tau_R \gg \tau_C$ with $\tau_C \approx 4\pi / \delta\epsilon$, $\tau_R := 1/R$ and $ R \approx \frac{4 \pi \, n 
 \,
\lambda^{2}}{\delta \epsilon}$. ($\tau_C$ corresponds to the autocorrelation function of $V$.) Or in other words there has to be a seperation of time scales between $\tau_R$ and $\tau_C$.
So, if the criteria \ref{eq-crit} are fulfilled, the expectation value dynamics are approximately described by $a(t) \approx a(0)\exp(-Rt)$ for the overwhelming majority of initial states \cite{bartsch2008,bartsch2009}, where the latter statement refers to the framework of ''typicality`` \cite{goldstein2006,popescu2006,reimann2007,bartsch2009}.
For a detailed discussion on the occurrence of exponential relaxation in this type of quantum models see \cite{bartsch2008}.
 An illustration of the graph of $a(t)$ is given in Fig. \ref{expcurve} and may also be found in
 \cite{bartsch2008,bartsch2009}. 
The decay rate is consistent with a pertinent projection operator approach or/and with Fermi's Golden Rule. 

Note that the exponential feature also depends on the choice of the observable $A$, i.e., for other observables the dynamical behavior could be completely different.

However, the dynamics as obtained by exact diagonalization show deviations from an exact exponential decay (``wiggles''), i.e., the complete dynamics may be viewed as a composition of a ``regular'' exponential part and some ``irregular'' looking part, the latter being perceived as fluctuations of the expectation value $a(t)$.

\textit{3. Entropy Definition and Numerical Analysis of the FT}
As stated at the very beginning of \cite{parrondo2009} there is a long and ongoing discussion about a microscopic non-equilibrium definition of entropy. Here we suggest an entropy $S(a)$ for which eventually a transient FT as defined in Eq. (\ref{fluc}) holds.
$S(a)$ is defined as $S(a) = N S(\rho_{\text{red}} (a))$ ($N=2n$ being the dimension of the full Hilbert space). $S(\rho_{\text{red}} (a))=- \text{Tr}\{ \rho_{\text{red}} \text{ln} \rho_{\text{red}} \}$ corresponds to a ``reduced'' von Neumann-entropy with 
\begin{equation}
\rho_{\text{red}}(a)=\frac{1}{N}((1+a)P_L+(1-a)P_R)\ ,
\end{equation}
i.e., $\rho_{\text{red}}$ is a ``reduced'' density matrix in the sense that it represents the maximum entropy state which is consistent with a given $a$. This of course is an implementation of Jaynes principle of maximum entropy as applied, e.g., in \cite{blankenbecler1985}.
Eventually, $S(a)$ exclusively depends on the expectation value of our only observed quantity $A$.
$S(a)$ becomes maximal for $a=0$, therefore $a=0$ is termed the ``equilibrium value'' in this context which is consistent with pertinent considerations within the framework of typicality \cite{goldstein2006,popescu2006,reimann2007,bartsch2009}.

The dynamics of $a(t)$ now yield concrete entropy dynamics, i.e., $S(a(t))$.
We first numerically investigate whether these entropy dynamics fulfill a FT in the sense of Eq. (\ref{fluc}).
To this end we determine a large number of time-averaged entropy productions ($\Sigma_{\tau}=(S(t+\tau)-S(t))/\tau$) for different $t$ and different trajectories, simply from numerically solving the Schr\"odinger equation.

\begin{figure}[htb]
\centering
\hspace{-1.0cm}
\includegraphics[width=6.8cm]{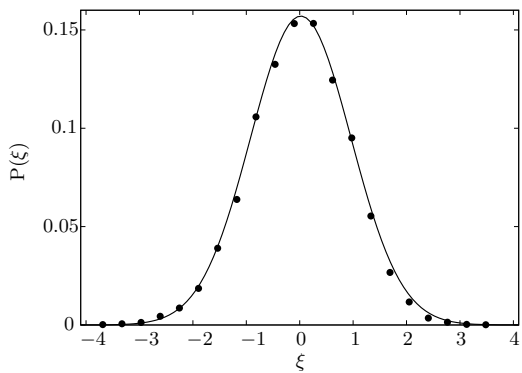}
\hspace{1.0cm}
\caption{Relative frequency distribution of normalized entropy productions $\xi := (\Sigma - \Sigma_0 )/ \sqrt{2 \Sigma_0 / \tau}$. The points represent the frequencies obtained from our numerics, the solid line is a Gaussian fit ($\mu \approx 0$, $\sigma \approx 0.94$, $\sigma^2 \approx 0.88$). System parameters: $N=6000$, $\delta\epsilon = 0.75$, $\lambda= 0.00013$.} 
\label{entprods}
\end{figure}

The mean entropy production $\Sigma_{0}$ is directly determined from the decay rate $R$.
Fig. \ref{entprods} shows a distribution of rescaled entropy productions, i.e., the mean value (as expected from the relaxation rate) and the variance (as expected by relation (\ref{gauss})) are ``scaled out'', for a system with parameters $N=6000, \delta\epsilon=0.75, \lambda=0.00013$. 
Note that we use only one realization of the Hamiltonian for these numerics, i.e., the ``perturbation'' matrix elements $v_{ij}$ are randomly chosen but fixed numbers.
The data is taken from  $1000$ trajectories departing from a set of initial states specified by a common $a(0)$ but chosen at random otherwise. From each trajectory we calculate entropy productions over $10$ segments which are located around an (average) expectation value of $a \approx 0.2 (0.21 - 0.16)$ at time $t\approx 820 - 1050$, i.e., there are $10000$ entropy productions in total.
The segment lengths $\tau$ are  chosen such that $\tau_C < \tau \ll \tau_R$ is fulfilled (i.e., $\tau \approx 70, \tau_C\approx 15, \tau_R = 1177$).

\begin{figure}[htb]
\centering
\hspace{-1.0cm}
\includegraphics[width=6.2cm]{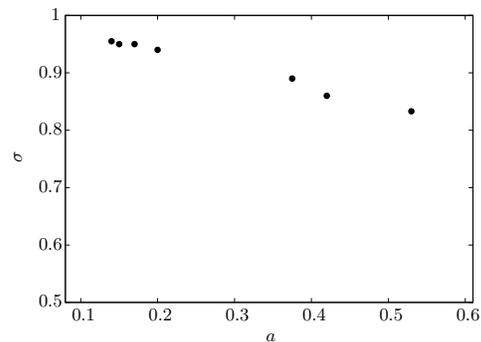}
\hspace{1.0cm}
\caption{Dependence of $\sigma$ on $a$. $\sigma$ decreases with larger deviation from equilibrium ($a=0$).  System parameters: $N=6000$, $\delta\epsilon = 0.75$, $\lambda= 0.00013$.} 
\label{varvsa}
\end{figure}

Obviously the distribution of the entropy productions is well described by a Gaussian (see Gaussian fit in Fig. \ref{entprods}). We find the mean and the width to be  $\mu \approx 0$ and be $\sigma = 0.94$, respectively. The FT would have been exactly fulfilled if we had found $\mu=0$ and $\sigma=1$.
That is, the variance is slightly smaller than postulated by the FT. Recall that the analyzed trajectory segments are picked at times corresponding to expectation values of $a\approx 0.2 (0.21 - 0.16)$ which means significantly off-equilibrium ($a=0$). Numerics (see Fig. \ref{varvsa}) suggest that the width gets closer to one with data taken closer to equilibrium for any model. The decrease of $\sigma$ for larger deviations from equilibrium cannot be fully explained within this framework. Further numerical investigations (which we omit here for brevity) indicate that the degree of the deviation from $\sigma=1$ when departing from equilibrium seems to depend on the specific model. Note that the model at hand has not been designed to be realistic in any detail, it has only been designed to be simple.

Based on these numerics we now formulate our first main result. We expect for models showing overall exponential relaxation of some quantum expectation value $a(t)$ and an entropy definition as $S(a)$ (see 3.),
the relative frequency of mean entropy productions over time steps which are larger than the correlation time but shorter than the relaxation time to fulfill a transient FT as given in  Eq. (\ref{fluc}). This is expected to hold within a regime around equilibrium the size of which may depend on the model.

\textit{4. Stochastic Description of Expectation Value Dynamics}
In the following we present some concepts of an (partially analytical) explanation for our main numerical results. The key idea is to show that the dynamics of an expectation value $a(t)$, as resulting from the fully deterministic Schr\"odinger Equation, may be described in terms of a time-discretized stochastic process of the following Ornstein-Uhlenbeck-type:
\begin{equation}
da_{i} = -R\, a_{i}\, \tau + \; \sqrt{2 R / N} \cdot dw_{i} 
\label{stoch}
\end{equation} 
Here the $dw_{i}$ are stochastic increments drawn from a standard Wiener process, i.e., the $dw_{i}$ are uncorrelated numbers drawn from a Gaussian distribution with zero mean, $\langle dw_{i}\rangle=0$, and normalized variance, $\langle dw_{i}dw_{j}\rangle=\delta_{ij}\tau$. Again the description is meant to hold for $\tau$ from a regime in between correlation and relaxation times. (Note that unlike in Nelson's approach \cite{nelson1966} we do not argue here that a quantum probability density may be found from a Fokker-Planck equation for a certain stochastic process, but that the evolution of a certain quantum expectation value has striking similarities with a sample path from some Ornstein-Uhlenbeck process.)

By hypothesizing Eq. (\ref{stoch}) and an entropy as given by a Taylor expansion of $S(a)$ (see 3.) up to the quadratic order around the equilibrium value $a=0$, i.e., $S(a)\approx N \text{ln}(N)- 1/2 \, N a^2$, 
one can analytically show that the probability distribution of entropy productions $P(\Sigma_{\tau})$ is Gaussian as well and fulfills the FT (\ref{fluc}). Applying Ito's lemma the entropy, which is a function of the expectation value, also becomes a stochastic process that is described by
\begin{equation}
dS_{i} = 2R (N \text{ln}(N)- S_{i}-\frac{1}{2})\, \tau - \; \sqrt{4 R (N \text{ln}(N)- S_{i})} \cdot dw_{i} \ .
\label{stochent}
\end{equation}
For large enough systems and large enough deviations from equilibrium, which is fulfilled in our setup, $N \text{ln}(N) - S_{i}$ is large compared to $1 / 2$. (Both $N \text{ln}(N)$ and $S_{i}$ become large for large $N$.) Consequently, the term $1/2$ in the first expression on the r.h.s. of (\ref{stochent}) may be neglected. Dividing by $\tau$ and further introducing the abbreviations $\Sigma_{i} := dS_{i}/\tau$ and $\Sigma_{0}:= 2R(N \text{ln}(N)- S_{i})$ one arrives at
\begin{equation}
\Sigma_{i}= \Sigma_{0}- \frac{\sqrt{2\Sigma_{0}}}{\tau} dw_{i} \ ,
\end{equation}
where $\Sigma_{i}$ corresponds to the entropy production over the time step $\tau$ and $\Sigma_{0}$ is the mean entropy production. That is, the distribution of entropy productions is Gaussian and the mean value and the variance obey the relation given in (\ref{gauss}), which means that the FT (\ref{fluc}) is fulfilled.

\begin{figure}[htb]
\centering
\hspace{-1.0cm}
\includegraphics[width=6.8cm]{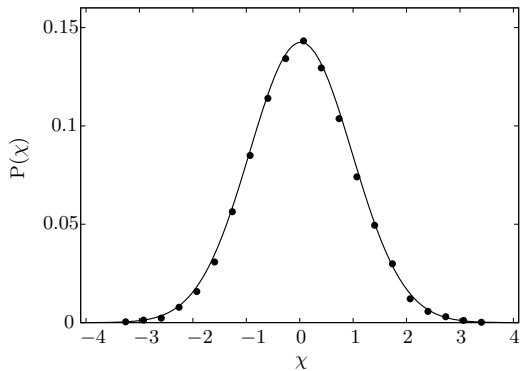}
\hspace{1.0cm}
\caption{Relative frequency distribution of normalized expectation value changes $\chi := D(\psi,t,\tau)/ \sqrt{2 R \tau / N}$. The points represent the frequencies obtained from our numerics, the solid line is a Gaussian fit ($\mu \approx 0$, $\sigma \approx 0.96$, $\sigma^2 \approx 0.92$). System parameters: $N=6000$, $\delta\epsilon = 0.75$, $\lambda= 0.00013$.}  
\label{increms}
\end{figure}

We now analyze whether the dynamics of $\langle \psi \vert A(t) \vert \psi \rangle$ according to the Schr\"odinger Equation are consistent with the crucial features of (\ref{stoch}). We especially regard the following points: (i) Is the overall behavior described by an exponential decay with the decay rate $R$? (ii) Is the distribution of the quantity
\begin{equation}
D(\psi,t,\tau) := \langle \psi \vert A(t+\tau) \vert \psi \rangle - \langle \psi \vert A(t) \vert \psi \rangle (1-R\tau) 
\label{ddef}
\end{equation}
Gaussian with zero mean and variance $\sigma^2 \approx 2 R \tau / N$? (Note that the deterministic $D(\psi,t,\tau)$ correspond to $\sqrt{2 R / N} \cdot dw_{i}$.) (iii) Are $D(\psi,t,\tau)$ and $D(\psi,t+\Delta t,\tau)$ ($\Delta t \geq \tau$) uncorrelated?

\begin{figure}[htb]
\centering
\hspace{-1.0cm}
\includegraphics[width=6.0cm]{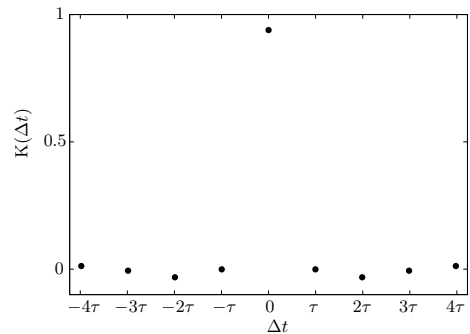}
\hspace{1.0cm}
\caption{Correlations of $\text{K}(\Delta t)=< D(\psi,t,\tau) D(\psi,t+\Delta t,\tau)>/ (2 R \tau / N)$. Each point is averaged over $1000$ trajectory segments, $\tau\approx 70$. System parameters: $N=6000$, $\delta\epsilon = 0.75$, $\lambda= 0.00013$.}  
\label{corrs}
\end{figure}

We start by investigating  those points numerically for our model. (i)  Here we simply resort to the results of  \cite{bartsch2008}, \cite{bartsch2009}. They give evidence that typically , up to small fluctuations, the  dynamics of $\langle \psi \vert A(t) \vert \psi \rangle$ is an exponential decay with a rate $R$ as given in 2. (ii) Numerics for a system of $N=6000, \delta\epsilon=0.75, \lambda=0.00013$ indicate that the distribution of the $D(\psi,t,\tau)\sqrt{N/ 2 R \tau }$ as generated from evaluating $10000$ segments along a single trajectory directly from the Schr\"odinger Equation, is approximately Gaussian with zero mean and $\sigma^2\approx 1$ (Fig. \ref{increms}) ($\tau \approx 70, \tau_C\approx 15, \tau_R = 1177$, again we use only one realization of $H$, i.e., the $v_{ij}$ are fixed).
This result appears to be robust against variation of the model parameters and time step lengths $\tau$ within the above regime. (iii) Moreover, Fig. \ref{corrs} shows that the correlations
$< D(\psi,t,\tau) D(\psi,t+\Delta t,\tau)>$, as obtained from the same data that has been exploited to clarify the previous issue (ii),  are nearly $0$ for the above chosen time step length $\tau$, such that the $D(\psi,t,\tau)$ can be considered as uncorrelated.

Thus, the numerics may suggest that for $\tau_C < \tau \ll \tau_R $ the deterministic $D(\psi,t,\tau)$ may be interpreted as stochastic variations $y_i$ and that a description by (\ref{stoch}) is valid in and also close to equilibrium.

In addition to the numerics we now provide some analytical considerations on the validity of a description as given by (\ref{stoch}), which also address the points (i)-(iii). Note that these considerations are independent of any details of the model, except for exponential approach to equilibrium. The statements are mathematically rigid, but on the level of ensemble arguments, e.g., it is shown below that the variance of the $D(\psi,t,\tau)$ is given by $2R\tau N^{-1}$ if the set (ensemble) of all possible states $\psi$ is considered. Strictly speaking this holds in itself no implication on the distribution of the $D(\psi,t,\tau)$ encountered along a single trajectory, i.e., with respect to a set of $t$. Such a connection only arises from some sort of ergodicity. But following the ideas of von Neumann \cite{goldstein2010}, this ergodicity may be expected here. The literature provides a formula for the ensemble or ``Hilbert space'' average (HA) of the product of two quantum ex
 pectation values. The ensemble includes all possible pure states from a finite dimensional Hilbert space \cite{gemmer2009}:
\begin{equation}
\text{HA}(\langle \psi \vert A\vert \psi \rangle \langle \psi \vert B\vert \psi \rangle) =(N(N+1))^{-1}\text{Tr} \{ AB\}
\label{hila}
\end{equation}
for $A,B$ traceless i.e., $ \text{Tr} \{ A\}=\text{Tr} \{ B\}=0$. Thus to address (i) we simply compute 
\begin{equation}
 \text{HA}(\langle \psi \vert A(t+\tau) \vert \psi \rangle \langle \psi \vert A(t)\vert \psi \rangle)=        \frac{ \text{Tr} \{ A(\tau)A \}   }{N(N+1)}  \ ,
\label{azerf}
\end{equation}
Since we focus here on models featuring exponential decay we have $ \text{Tr} \{ A(\tau)A \}\propto e^{-R\tau }\approx 1-R\tau $. This justifies the usage of $R$ in (\ref{stoch}). To address (ii) we calculate the  Hilbert space average  of  $D^2(\psi,t,\tau)$ from (\ref{ddef}), which should be a measure for the fluctuations. 
By means of Eqs. (\ref{hila}) and (\ref{azerf}) one obtains
\begin{eqnarray}
\text{HA}(D^2 (\psi,t,\tau)))&=& (\text{Tr}\{ A^2 (t+\tau)\} \\ \nonumber
&-&2(1-R\tau) \text{Tr}\{ A(t+\tau)A(t)\} \\ \nonumber
&+& (1-R\tau)^2 \text{Tr} \{ A^2 (t)\})(N(N+1))^{-1} \ .
\end{eqnarray}
Assuming the above exponential decay of the correlation function one obtains to linear order in $\tau$
\begin{equation}
\text{HA}(D^2 (\psi,t,\tau)) \approx 2R\tau N^{-1} \ ,
\label{dvar}
\end{equation}
i.e., the fluctuations become small for large $N$ and are proportional to the relaxation rate.
The result of (\ref{dvar}) is exactly what we assumed as the width of the stochastic fluctuations in Eq. (\ref{stoch}). Regarding (iii), one finds by realizing that
\begin{equation}
D(\psi,t,2\tau)=D(\psi,t,\tau)+D(\psi,t+\tau,\tau) \ ,
\end{equation}
and exploiting (\ref{dvar}) that the HA of the correlations between subsequent trajectory segments vanishes
\begin{equation}
\text{HA}(D(\psi,t,\tau)D(\psi,t+\tau,\tau))=0 \ .
\label{HAcorr}
\end{equation}
(\ref{HAcorr}) may be straightforwardly extended to non-subsequent trajectory segments.

To repeat, the above HA calculations address averages over all possible states in Hilbert space and therefore yield results only on the average behavior of all possible trajectories, but not directly for individual trajectories. Since Eq. (\ref{stoch}) is certainly meant to describe individual trajectories, the HA results cannot strictly prove the corresponding features of (\ref{stoch}). Nevertheless, the HA results are at least consistent with (\ref{stoch}), which is not self-evident, and may 
in so far strengthen the trust in the validity of a description as given by (\ref{stoch}).

\textit{5. Conclusion}
In this Letter we present an analysis of temporal fluctuations in closed quantum systems featuring exponential relaxation of some expectation value. Numerics indicate that the dynamics of an entropy-like function are correctly described by a FT. This feature directly arises from Schr\"odinger-type dynamics, i.e., no assumption of, e.g., a Quantum Master Equation is involved. This result appears to be based on the finding that the underlying expectation value relaxation dynamics may be interpreted in terms of a stochastic process.
In how far these results may be extended to quantities, which are better accessible by experiments, like, e.g., work, is an open question, particularly because work is not a quantum observable \cite{talkner2007}.

We thank C. Maes, A. Engel, H. Niemeyer and P. Maas for fruitful discussions.


\end{document}